\documentclass{article}
\usepackage{amsfonts,amssymb,amsmath}
\usepackage{color, xcolor}
\usepackage{listings}
\usepackage{graphicx}
\usepackage{url}
\usepackage{ulem}

\newcommand{\Imm}{\mathrm{Im}}

\newcommand {\U}{\mathrm{U}}

\newcommand {\SO}{\mathrm{SO}}
\newcommand {\SU}{\mathrm{SU}}
\newcommand {\Sp}{\mathrm{Sp}}
\newcommand {\sso}{\mathfrak{so}}
\newcommand {\gl}{\mathfrak{gl}}
\renewcommand {\u}{\mathfrak{u}}
\newcommand {\su}{\mathfrak{su}}
\renewcommand {\sl}{\mathfrak{sl}}

\newcommand{\Sym}{\mathrm{Sym}}
\newcommand {\Ker}{\mathrm{Ker}}

\newcommand{\beq}{\begin{equation}}
\newcommand{\eeq}{\end{equation}}
\newcommand{\beqr}{\begin{eqnarray}}
\newcommand{\eeqr}{\end{eqnarray}}

\stepcounter{equation}

\begin{document}
\title {Homology of Lie algebra of supersymmetries}
\author{M. V. Movshev\\Stony Brook University\\Stony Brook, NY 11794-3651, USA
\\ A. Schwarz\\ Department of Mathematics\\ University of
California \\ Davis, CA 95616, USA, \\Renjun Xu \\Department of Physics, University of California \\ Davis, CA 95616, USA }

\date{\today}
\maketitle

Abstract { We study the homology and cohomology groups of super Lie algebra of supersymmetries  and of super Poincare algebra. We discuss in detail the calculation in dimensions D=10 and D=6. Our  methods can be applied  to extended supersymmetry algebra and to other dimensions.}

\section {Introduction}

In present paper we will analyze homology and cohomology groups of the super Lie algebra of supersymmetries and of super Poincare Lie algebra. We came to this problem studying supersymmetric
deformations of maximally supersymmetric gauge theories ~\cite {MS}; however, this problem arises also in different situations, in particular, in supergravity~\cite{Supergravity}.  In low dimensions it was studied in~\cite{F.Brandt}

Let us recall the definition of Lie algebra cohomology. We start with super Lie algebra $\cal G$ with generators $e_A$ and structure constants $f_{AB}^K$. We introduce ghost variables $C^A$ with
parity opposite to the parity of generators $e_A$ and consider the algebra $E$ of polynomial functions of these variables. (In more invariant way we can say that $E$ consists of polynomial functions
on linear superspace $\Pi \cal G$.) We define a derivation $d$ on $E$ by the formula
   $d=\frac{1}{2}f^K_{AB}C^AC^B\frac{\partial}{\partial C^K}.$

This operator is a differential (i.e. it changes the parity and obeys $d^2=0.$) We define the cohomology of $\cal G$ using this differential:
$$H^{\bullet}({\cal G})= \Ker d/\Imm d.$$
The definition of homology of $\cal G$ is dual to the definition of cohomology: instead of $E$ we consider its dual space $E^*$ that can be considered as the space of functions of dual ghost variables
$c_A$; the differential $\partial$ on $E^*$ is defined as an operator adjoint to $d$. The homology $H_{\bullet}(\cal G)$ is dual to the cohomology $H^{\bullet}(\cal G)$.

Notice that we can multiply cohomology
classes, i.e. $H^{\bullet}(\cal G)$ is an algebra.

The super Lie algebra of supersymmetries has odd generators $e_{\alpha}$ and even generators $P_m$ ; the only non-trivial  commutation relation is
$$[e_{\alpha}, e_{\beta}]_+=\Gamma _{\alpha \beta}^m P_m.$$
The coefficients in this relation are Dirac Gamma matrices. The space $E$ used in the definition of cohomology  (cochain complex) consists here of polynomial functions of even ghost variables $t^{\alpha}$ and odd ghost
variables $ c^m$; the differential has the form
$$d=\frac{1}{2}\Gamma _{\alpha \beta}^mt^{\alpha}t^{\beta}\frac{\partial}{\partial c^m}.$$
The space $E$ is double-graded (one can consider the degree with respect to $t^{\alpha}$ and the degree with respect to $c^m$). In more invariant form we can say that {\footnote {We use the notation
$\Sym^m$ for symmetric tensor power and the notation $\Lambda ^n $ for exterior power}}
$$E=\sum \Sym^m S\otimes \Lambda ^n V$$
where $S$ stands for spinorial representation of orthogonal group, $V$ denotes vector representation of this group and Gamma-matrices specify an intertwiner $V\to \Sym^2 S$. The differential $d$ maps
$\Sym^m S\otimes \Lambda ^n V$ into $\Sym^{m+2} S\otimes \Lambda ^{n-1}V$. The description above can be applied to any dimension and to any signature of the metric used in the definition of orthogonal
group, however,  the choice of spinorial representation is different in different dimensions. {\footnote {Recall that orthogonal group $\SO(2n)$ has two irreducible two-valued complex representations called semi-spin representations (left spinors and right spinors), the orthogonal group $\SO(2n+1)$ has one irreducible two-valued complex spin representation. One says that a real representation is spinorial if after extension of scalars to $\mathbb{C}$ it becomes a sum of spin or semi-spin representations. (We follow the terminology of \cite {Del}.)}}  The group $\SO(n)$
can be considered as a (subgroup) of the group of automorphisms of supersymmetry Lie algebra and therefore it acts on its cohomology.

We will  start with ten-dimensional case; in this case the  spinorial representation
should be considered as one of  two irreducible two-valued 16-dimensional representations of $\SO(10)$ (the spinors are Majorana-Weyl spinors).  We will work with complex representations and complex
group $\SO(10)$; this  does not change the cohomology.

 The double grading on $E$ induces double grading on cohomology. However, instead of the degrees $m$ and $n$ it is more convenient to use the degrees
$k=m+2n$ and $n$ because the differential preserves $k$ and therefore the problem of calculation of cohomology can be solved for every $k$ separately. It important to notice that the differential
commutes with multiplication by a polynomial depending on $t^{\alpha}$, therefore the cohomology is a module over the polynomial ring ${\bf{C}}[t^1,..., t^{\alpha},...].$ (Moreover, it is an algebra
over this ring.) The cohomology is infinite-dimensional as a vector space, but it has a finite number of generators as a ${\bf{C}}[t^1,..., t^{\alpha},...]$-module (this follows from the fact that the
polynomial ring is noetherian). One of the most important problems is the description of these generators.

The action of orthogonal group on $E$ commutes with the differential, therefore the orthogonal group acts on cohomology. (This action is two-valued, hence it would be more precise to talk about the
action of the spinor group or about the action of the corresponding Lie algebra).

We will describe now the cohomology of the Lie algebra of  supersymmetries in ten-dimensional case as representations of the Lie algebra $\sso_{10}$. As usual the representations are labeled by their
highest weight. The vector representation $V$ has the highest weight $[1,0,0,0,0]$, the irreducible spinor representations have highest weights $[0,0,0,0,1]$,$[0,0,0,1,0]$; we assume that the highest weight of $S$ is $[0,0,0,0,1]$. The description
of graded component of cohomology group with gradings $k=m+2n$ and $n$ is given by the formulas for $H^{k,n}$ (for $n\geq 6$, $H^{k,n}$ vanishes)
\beqr
H^{k,0}&=& [0,0,0,0,k] \label{Hk,0}\\
H^{k,1}&=& [0,0,0,1,k-3] \\
H^{k,2}&=& [0,0,1,0,k-6] \\
H^{k,3}&=& [0,1,0,0,k-8] \\
H^{k,4}&=& [1,0,0,0,k-10] \\
H^{k,5}&=& [0,0,0,0,k-12] \label{Hk,5}
\eeqr
The only special case is when $k=4$, there is one additional term, a scalar, for $H^{4,1}$.
\beq
H^{4,1}=[0,0,0,0,0]+[0,0,0,1,1] \label{H4,1}
\eeq

The cohomology considered as ${\bf{C}}[t^1,..., t^{\alpha},...]$-module is generated by $H^{1,0}$, $H^{3,1}$, $H^{6,2}$, $H^{8,3}$, $H^{10,4}$, $H^{12,5}$.

Let us discuss  shortly the Lie algebra of supersymmetries in other dimensions (see \cite {Del} for more detail). We will start with
the case of the space with Minkowski signature (the case of orthogonal group $\SO(1,n-1)$).
In this case for an irreducible spinorial representation $S$ there a unique (up to a factor) intertwiner
$V\to \Sym^2S$; we use this intertwiner in the definition of the Lie algebra of supersymmetries. Real representations are classified according the structure of their algebra of endomorphisms: if this algebra is isomorphic to $\mathbb{C}$ one says that the real representation is complex, if the algebra is isomorphic to quaternions one says that the representation is quaternionic. Irreducible spinorial representations in Minkowski case are complex  for $n=8k$ and $n=8k+4$, they are quaternionic for $n=8k+5,8k+6,8k+7$; correspondingly their automorphism groups contain $\U(1)$
and $\Sp(1)=\SU(2)$.  In the complex case the cohomology can de considered as  a representation of the group $\SO(n)\times \U(1)$ (or, more precisely of the Lie algebra $\sso_{10}\times \u_1$), in the quaternionic case we obtain a representation of the group $\SO(n)\times \SU(2)$ (of the Lie algebra $\sso_n\times \su_2$). It will be convenient for us to complexify the Lie algebra of supersymmetries; the complexification does not change the cohomology. The cohomology can be considered as a representation of the group of automorphisms of the supersymmetry Lie algebra, of  Lie algebra of this group or of the complexified Lie algebra. The complexified Lie algebra is  $\sso_n\times \gl_1$ if $S$ is a complex representation and  $\sso_n\times \sl(2)$ in quaternionic case. It is isomorphic to $\sso_n$ if the algebra of endomorphisms  of $S$ is isomorphic to $\mathbb{R}.$  (We abuse notations denoting the complexified orthogonal Lie algebra in the same way as its real counterpart.)

One can consider also $N$-extended supersymmetry Lie algebra. In the case of Minkowski signature this means that we  should start with reducible spinorial representation $S_N$(direct sum of $N$ copies of irreducible spinorial representation $S$). Taking $N$ copies of the intertwiner  $V\to \Sym^2S$ we obtain an intertwiner $V\to \Sym^2S_N$. We define the $N$-extended supersymmetry Lie algebra by means of this intertwiner. The Lie algebra acting on its cohomology acquires an additional factor $\u_N$ ( or $\gl_N$ if we work with complex Lie algebras).

Let us consider, for example the six-dimensional case. In this case there are two irreducible spinorial representations, after extension of scalars to $\mathbb{C}$ each of these representations becomes a direct sum $S=S_0+S_0=S_0\times T$ of two equivalent semi-spin representations. (Here $T$ stands for two-dimensional space. )The intertwiner $V\to \Sym^2S$ can be obtained as a tensor product of maps $V\to \Lambda ^2S_0$ and $\mathbb{C}\to \Lambda ^2T$.

Now we will describe the cohomology of the Lie algebra of  supersymmetries in six-dimensional case as representations of the Lie algebra $\sso(6)\times \sl_2$. The vector representation $V$ of $\sso(6)$ has the highest weight $[1,0,0]$, the irreducible spinor representations have highest weights $[0,0,1]$, $[0,1,0]$; we assume that the highest weight of $S_0$ is $[0,0,1]$. As a representation $\sso(6)\times \sl_2$ the representation $V$ has the weight $[1,0,0,0]$ and the representation $S=S_0\times T$ has the weight $[0,0,1,1]$. The description
of graded component of cohomology group with gradings $k=m+2n$ and $n$ is given by the formulas of $H^{k,n}$ (for $n\geq 4$, $H^{k,n}$ vanishes)
\beqr
H^{k,0}&=& [0,0,k,k] \label{Hk6D,0}\\
H^{k,1}&=& [0,1,k-3,k-2] \\
H^{k,2}&=& [1,0,k-6,k-4] \\
H^{k,3}&=& [0,0,k-8,k-6] \label{Hk6D,3}
\eeqr
The only special case is when $k=4$, there is one additional term, a scalar, for $H^{4,1}$.
\beq
H^{4,1}=[0,0,0,0]+[0,1,1,2] \label{H46D,1}
\eeq

The cohomology considered as ${\bf{C}}[t^1,..., t^{\alpha},...]$-module is generated by $H^{1,0}$, $H^{3,1}$, $H^{6,2}$, $H^{8,3}$.

There are different ways to perform these calculations. In this paper we describe the most elementary way. We used the program LiE  ~\cite{LiEcode} to decompose $ \Sym^m S\otimes \Lambda ^n V$
into irreducible representation of  automorphism Lie algebra for small $k=m+2n$. We used the result to guess the general answer ; we check it by means of Weyl dimension formula. Due to Schur's lemma one can consider every irreducible representation separately.  Assuming that the
kernel of the differential is as small as possible ("principle of maximal propagation") we calculate the cohomology. We justify this calculation using the fact that the differential  commutes with multiplication by a polynomial ghost variables $t^{\alpha}$ and therefore the multiplication by such a polynomial transform a boundary (an element in the image of differential) into a boundary . We write down explicitly the decomposition of $ \Sym^m S\otimes \Lambda ^n V$
into irreducible representation of  automorphism Lie algebra overlining the images of the differential (the boundaries) and underlining the terms mapped to the boundaries by the differential; the remaining terms give the decomposition of cohomology.

\section{Calculations for D=10}

To calculate the cohomology we decompose each graded component $E^{kn}=\Sym^{k-2n}S\otimes \Lambda ^n V$ of $E$ into direct sum of irreducible representations.

For $D=10$ spacetime, we have the cochain complex
\beq
\begin{split}
0& \xleftarrow{d_0} \Sym^k(S) \xleftarrow{d_1} \Sym^{k-2}(S)\otimes V \xleftarrow{d_2} \Sym^{k-4}(S)\otimes \wedge^2 V \\
&\xleftarrow{d_3} \Sym^{k-6}(S)\otimes \wedge^3 V \xleftarrow{d_4} \Sym^{k-8}(S)\otimes \wedge^4V \xleftarrow{d_5}\Sym^{k-10}(S)\otimes \wedge^5V \\
&\xleftarrow{d_6}\Sym^{k-12}(S)\otimes \wedge^6V  \xleftarrow{d_7}\Sym^{k-14}(S)\otimes \wedge^7V \xleftarrow{d_8}\Sym^{k-16}(S)\otimes \wedge^8V \\
&\xleftarrow{d_9}\Sym^{k-18}(S)\otimes \wedge^9V \xleftarrow{d_{10}}\Sym^{k-20}(S)\otimes \wedge^{10}V \xleftarrow{d_{11}} 0 \label{complex.10D}
\end{split}
\eeq
where for $\Sym^m(S)\otimes \wedge^n(V)$, a grading degree defined by $k=m+2n$ is invariant upon cohomological differential $d$. All components of this complex can be regarded as representations of $\sso(10)$. We have
\beq
\begin{split}
&S=[0,0,0,0,1] \text{ (choosen) or } [0,0,0,1,0], \quad V=[1, 0,0,0,0] \\
&\wedge^2V=[0,1,0,0,0], \quad \wedge^3V=[0,0,1,0,0],    \\
&\wedge^4V=[0,0,0,1,1], \quad \wedge^5V=[0,0,0,0,2]\oplus [0,0,0,2,0],    \\
&\wedge^6V=\wedge^4V, \quad \wedge^7V=\wedge^3V, \quad \wedge^8V=\wedge^2;V, \quad \wedge^9V=V, \quad \wedge^{10}V=[0,0,0,0,0],   \\
\end{split}
\eeq

For $\Sym^m S\otimes \wedge^n V$, where $m\geq 1$,
\beq
\Sym^k(S)=
\overset{\left[k/2\right]}{\underset{i=1}{\oplus}}{\color{blue}\underline{[i,0,0,0,k-2i]}}\oplus [0,0,0,0,k],  \label{SymkS_10D}
\eeq
\beq
\begin{split}
\Sym^{k-2}(S)\otimes V&=
 \overset{\left[k/2\right]}{\underset{i=1}{\oplus}}{\color{red}\overline{[i,0,0,0,k-2i]}} \overset{\left[(k-4)/2\right]}{\underset{i=0}{\oplus}}{\color{blue}\underline{[i,0,0,0,k-4-2i]}} \\
 &\overset{\left[(k-3)/2\right]}{\underset{i=1}{\oplus}}{\color{blue}\underline{[i,0,0,1,k-3-2i]}} \oplus [0,0,0,1,k-3] \\ &\overset{\left[(k-4)/2\right]}{\underset{i=0}{\oplus}}{\color{blue}\underline{[i,1,0,0,k-4-2i]}},
\end{split}
\eeq
\beq
\begin{split}
\Sym^{k-4}(S)\otimes \wedge^2V &=
 \overset{\left[(k-4)/2\right]}{\underset{i=0}{\oplus}} {\color{red}\overline{[i,0,0,0,k-4-2i]}}
 \overset{\left[(k-5)/2\right]}{\underset{i=1}{\oplus}}{\color{blue}\underline{[i,0,0,0,k-4-2i]}} \\
&\overset{\left[(k-3)/2\right]}{\underset{i=1}{\oplus}} {\color{red}\overline{[i,0,0,1,k-3-2i]}}
 \overset{\left[(k-7)/2\right]}{\underset{i=0}{\oplus}}{\color{blue}\underline{[i,0,0,1,k-7-2i]}} \\
&\overset{\left[(k-4)/2\right]}{\underset{i=0}{\oplus}} {\color{red}\overline{[i,1,0,0,k-4-2i]}}
 \overset{\left[(k-8)/2\right]}{\underset{i=0}{\oplus}}{\color{blue}\underline{[i,1,0,0,k-8-2i]}} \\
&\overset{\left[(k-6)/2\right]}{\underset{i=1}{\oplus}}{\color{blue}\underline{2[i,0,1,0,k-6-2i]}} \oplus {\color{blue}\underline{[0,0,1,0,k-6]}} \\
&\oplus [0,0,1,0,k-6]
 \overset{\left[(k-7)/2\right]}{\underset{i=0}{\oplus}}{\color{blue}\underline{[i,1,0,1,k-7-2i]}},
\end{split}
\eeq
\beq
\begin{split}
\Sym^{k-6}(S)\otimes \wedge^3V &=
\overset{\left[(k-9)/2\right]}{\underset{i=0}{\oplus}}{\color{blue}\underline{[i,0,0,0,k-8-2i]}}
 \overset{\left[(k-5)/2\right]}{\underset{i=1}{\oplus}} {\color{red}\overline{[i,0,0,0,k-4-2i]}} \\
&\overset{\left[(k-7)/2\right]}{\underset{i=0}{\oplus}} {\color{red}\overline{[i,0,0,1,k-7-2i]}}
 \overset{\left[(k-8)/2\right]}{\underset{i=1}{\oplus}}{\color{blue}\underline{2[i,0,0,1,k-7-2i]}}\\
&\oplus {\color{blue}\underline{[0,0,0,1,k-7]}} {\underset{k-odd}{\oplus}} {\color{blue}\underline{[\frac{k-7}{2},0,0,1,0]}} \\
&\overset{\left[(k-10)/2\right]}{\underset{i=0}{\oplus}}{\color{blue}\underline{2[i,0,1,0,k-10-2i]}}
 \oplus {\color{red}\overline{[0,0,1,0,k-6]}} \\
&\overset{\left[(k-6)/2\right]}{\underset{i=1}{\oplus}}{\color{red}\overline{2[i,0,1,0,k-6-2i]}} \overset{\left[(k-9)/2\right]}{\underset{i=0}{\oplus}}{\color{blue}\underline{[i,0,1,1,k-9-2i]}}\\
&\oplus {\color{blue}\underline{[0,1,0,0,k-8]}} \oplus [0,1,0,0,k-8] \\
&\overset{\left[(k-9)/2\right]}{\underset{i=1}{\oplus}}{\color{blue}\underline{2[i,1,0,0,k-8-2i]}}
 \overset{\left[(k-8)/2\right]}{\underset{i=0}{\oplus}} {\color{red}\overline{[i,1,0,0,k-8-2i]}}\\
& {\underset{k-even}{\oplus}} {\color{blue}\underline{[\frac{k-8}{2},1,0,0,0]}}
\overset{\left[\frac{k-7}{2}\right]}{\underset{i=0}{\oplus}} {\color{red}\overline{[i,1,0,1,k-7-2i]}}\\
&\overset{\left[\frac{k-11}{2}\right]}{\underset{i=0}{\oplus}} {\color{blue}\underline{[i,1,0,1,k-11-2i]}}
 \overset{\left[\frac{k-10}{2}\right]}{\underset{i=0}{\oplus}}{\color{blue}\underline{[i,1,1,0,k-10-2i]}},
\end{split}
\eeq
\beq
\begin{split}
\Sym^{k-8}(S)\otimes \wedge^4V &=
{\color{blue}\underline{[0,0,0,0,k-8]}} \oplus [1,0,0,0,k-10] \oplus {\color{blue}\underline{2[1,0,0,0,k-10]}} \\
&\overset{\left[(k-9)/2\right]}{\underset{i=0}{\oplus}} {\color{red}\overline{[i,0,0,0,k-8-2i]}}
 \overset{\left[(k-10)/2\right]}{\underset{i=2}{\oplus}}{\color{blue}\underline{3[i,0,0,0,k-8-2i]}} \\
&{\underset{k-odd}{\oplus}} {\color{blue}\underline{2[\frac{k-9}{2},0,0,0,1]}} {\underset{k-even}{\oplus}}{\color{blue}\underline{[\frac{k-8}{2},0,0,0,0]}}\\
& \oplus {\color{red}\overline{[0,0,0,1,k-7]}} \overset{\left[(k-8)/2\right]}{\underset{i=1}{\oplus}}{\color{red}\overline{2[i,0,0,1,k-7-2i]}} \\
 &{\underset{k-odd}{\oplus}} {\color{red}\overline{[\frac{k-7}{2},0,0,1,0]}}
  \overset{\left[(k-12)/2\right]}{\underset{i=0}{\oplus}}{\color{blue}\underline{2[i,0,0,1,k-11-2i]}} \\
 &{\underset{k-odd}{\oplus}}{\color{blue}\underline{[\frac{k-11}{2},0,0,1,0]}}
  \overset{\left[(k-10)/2\right]}{\underset{i=0}{\oplus}}{\color{blue}\underline{[i,0,0,2,k-10-2i]}} \\
 &\overset{\left[(k-10)/2\right]}{\underset{i=0}{\oplus}}{\color{red}\overline{2[i,0,1,0,k-10-2i]}}
  \oplus {\color{blue}\underline{[0,0,1,0,k-10]}} \\
 & \overset{\left[(k-11)/2\right]}{\underset{i=1}{\oplus}}{\color{blue}\underline{2[i,0,1,0,k-10-2i]}} {\underset{k-even}{\oplus}}{\color{blue}\underline{[\frac{k-10}{2},0,1,0,0]}}\\
 & \overset{\left[(k-12)/2\right]}{\underset{i=0}{\oplus}}{\color{blue}\underline{[i,0,2,0,k-12-2i]}} \overset{\left[(k-9)/2\right]}{\underset{i=0}{\oplus}} {\color{red}\overline{[i,0,1,1,k-9-2i]}}\\
 & \overset{\left[(k-13)/2\right]}{\underset{i=0}{\oplus}}{\color{blue}\underline{[i,0,1,1,k-13-2i]}} \oplus {\color{red}\overline{[0,1,0,0,k-8]}}\\
 & \overset{\left[(k-9)/2\right]}{\underset{i=1}{\oplus}}{\color{red}\overline{2[i,1,0,0,k-8-2i]}}
   {\underset{k-even}{\oplus}} {\color{red}\overline{[[\frac{k-8}{2}],1,0,0,0]}} \\
 &\overset{\left[(k-13)/2\right]}{\underset{i=0}{\oplus}}{\color{blue}\underline{2[i,1,0,0,k-12-2i]}}
   {\underset{k-even}{\oplus}}{\color{blue}\underline{[\frac{k-12}{2},1,0,0,0]}} \\
 &\overset{\left[(k-12)/2\right]}{\underset{i=0}{\oplus}}{\color{blue}\underline{[i,2,0,0,k-12-2i]}}\\
 &\overset{\left[(k-11)/2\right]}{\underset{i=0}{\oplus}}{\color{red}\overline{[i,1,0,1,k-11-2i]}}
  \overset{\left[(k-11)/2\right]}{\underset{i=0}{\oplus}}{\color{blue}\underline{[i,1,0,1,k-11-2i]}} \\
 &\overset{\left[(k-10)/2\right]}{\underset{i=0}{\oplus}} {\color{red}\overline{[i,1,1,0,k-10-2i]}}
  \overset{\left[(k-14)/2\right]}{\underset{i=0}{\oplus}}{\color{blue}\underline{[i,1,1,0,k-14-2i]}},  
\end{split}
\eeq
\beq
\begin{split}
\Sym^{k-10}(S)\otimes \wedge^5V &=
{\color{red}\overline{\left[0,0,0,0,k-8\right]}} \oplus {\color{red}\overline{2[1,0,0,0,k-10]}} \overset{\left[(k-10)/2\right]}{\underset{i=2}{\oplus}} {\color{red}\overline{3[i,0,0,0,k-8-2i]}} \\
&{\underset{k-odd}{\oplus}} {\color{red}\overline{2[\frac{k-9}{2},0,0,0,1]}} {\underset{k-even}{\oplus}} {\color{red}\overline{[\frac{k-8}{2},0,0,0,0]}} \\
&\oplus [0,0,0,0,k-12] \overset{\left[(k-14)/2\right]}{\underset{i=1}{\oplus}}{\color{blue}\underline{3[i,0,0,0,k-12-2i]}} \\
& {\underset{k-odd}{\oplus}}{\color{blue}\underline{2[\frac{k-13}{2},0,0,0,1]}} \oplus {\color{blue}\underline{2[0,0,0,0,k-12]}} \\
&{\underset{k-even}{\oplus}}{\color{blue}\underline{[\frac{k-12}{2},0,0,0,0]}} \oplus {\color{blue}\underline{[0,0,0,1,k-11]}} \\
&\overset{\left[(k-12)/2\right]}{\underset{i=0}{\oplus}}{\color{red}\overline{2[i,0,0,1,k-11-2i]}} \overset{\left[(k-12)/2\right]}{\underset{i=1}{\oplus}}{\color{blue}\underline{2[i,0,0,1,k-11-2i]}} \\
&{\underset{k-odd}{\oplus}}{\color{red}\overline{[\frac{k-11}{2},0,0,1,0]}} {\underset{k-odd}{\oplus}}{\color{blue}\underline{[\frac{k-11}{2},0,0,1,0]}}\\
&\overset{\left[(k-10)/2\right]}{\underset{i=0}{\oplus}}{\color{red}\overline{[i,0,0,2,k-10-2i]}} \overset{\left[(k-14)/2\right]}{\underset{i=0}{\oplus}}{\color{blue}\underline{[i,0,0,2,k-14-2i]}} \\
&\oplus {\color{red}\overline{[0,0,1,0,k-10]}} \overset{\left[(k-11)/2\right]}{\underset{i=1}{\oplus}}{\color{red}\overline{2[i,0,1,0,k-10-2i]}} \\
&\oplus {\underset{k-even}{\oplus}}{\color{red}\overline{[\frac{k-10}{2},0,1,0,0]}} \overset{\left[(k-15)/2\right]}{\underset{i=0}{\oplus}}{\color{blue}\underline{2[i,0,1,0,k-14-2i]}} \\ &{\underset{k-even}{\oplus}}{\color{blue}\underline{[\frac{k-14}{2},0,1,0,0]}} \overset{\left[(k-12)/2\right]}{\underset{i=0}{\oplus}}{\color{red}\overline{[i,0,2,0,k-12-2i]}}\\
&\overset{\left[(k-16)/2\right]} {\underset{i=0}{\oplus}}{\color{blue}\underline{[i,0,2,0,k-16-2i]}} \overset{\left[(k-13)/2\right]}{\underset{i=0}{\oplus}}{\color{red}\overline{[i,0,1,1,k-13-2i]}} \\ &\overset{\left[(k-13)/2\right]}{\underset{i=0}{\oplus}}{\color{blue}\underline{[i,0,1,1,k-13-2i]}} \oplus {\color{blue}\underline{[0,1,0,0,k-12]}} \\
&\overset{\left[(k-13)/2\right]}{\underset{i=0}{\oplus}}{\color{red}\overline{2[i,1,0,0,k-12-2i]}} \overset{\left[(k-13)/2\right]}{\underset{i=1}{\oplus}}{\color{blue}\underline{2[i,1,0,0,k-12-2i]}}\\ &{\underset{k-even}{\oplus}}{\color{red}\overline{[\frac{k-12}{2},1,0,0,0]}} {\underset{k-even}{\oplus}}{\color{blue}\underline{[\frac{k-12}{2},1,0,0,0]}}  \\
&\overset{\left[(k-12)/2\right]}{\underset{i=0}{\oplus}}{\color{red}\overline{[i,2,0,0,k-12-2i]}} \overset{\left[(k-16)/2\right]}{\underset{i=0}{\oplus}}{\color{blue}\underline{[i,2,0,0,k-16-2i]}} \\
&\overset{\left[(k-15)/2\right]}{\underset{i=0}{\oplus}}{\color{blue}\underline{[i,1,0,1,k-15-2i]}} \overset{\left[(k-11)/2\right]}{\underset{i=0}{\oplus}}{\color{red}\overline{[i,1,0,1,k-11-2i]}} \\
&\overset{\left[(k-14)/2\right]}{\underset{i=0}{\oplus}}{\color{red}\overline{[i,1,1,0,k-14-2i]}} \overset{\left[(k-14)/2\right]}{\underset{i=0}{\oplus}}{\color{blue}\underline{[i,1,1,0,k-14-2i]}} \label{SymkS5V_10D}
\end{split}
\eeq
\beq
\begin{split}
\Sym^{k-12}(S)\otimes \wedge^6V &=
{\color{red}\overline{2[0,0,0,0,k-12]}} \overset{\left[(k-14)/2\right]}{\underset{i=1}{\oplus}}{\color{red}\overline{3[i,0,0,0,k-12-2i]}} \\
&{\underset{k-odd}{\oplus}} {\color{red}\overline{2[\frac{k-13}{2},0,0,0,1]}} \overset{\left[(k-13)/2\right]}{\underset{i=1}{\oplus}}{\color{blue}\underline{[i,0,0,0,k-12-2i]}}\\
 &{\underset{k-even}{\oplus}}{\color{red}\overline{[\frac{k-12}{2},0,0,0,0]}} \oplus {\color{red}\overline{[0,0,0,1,k-11]}} \overset{\left[(k-12)/2\right]}{\underset{i=1}{\oplus}}{\color{red}\overline{2[i,0,0,1,k-11-2i]}} \\
 &{\underset{k-odd}{\oplus}}{\color{red}\overline{[\frac{k-11}{2},0,0,1,0]}} \overset{\left[(k-16)/2\right]}{\underset{i=0}{\oplus}}{\color{blue}\underline{2[i,0,0,1,k-15-2i]}} \\
 &{\underset{k-odd}{\oplus}}{\color{blue}\underline{[\frac{k-15}{2},0,0,1,0]}} \overset{\left[(k-14)/2\right]}{\underset{i=0}{\oplus}}{\color{red}\overline{[i,0,0,2,k-14-2i]}} \\
 &\overset{\left[(k-15)/2\right]}{\underset{i=0}{\oplus}}{\color{red}\overline{2[i,0,1,0,k-14-2i]}} {\underset{k-even}{\oplus}}{\color{red}\overline{[\frac{k-14}{2},0,1,0,0]}}\\
 &\oplus {\color{blue}\underline{[0,0,1,0,k-14]}} \overset{\left[(k-14)/2\right]}{\underset{i=1}{\oplus}}{\color{blue}\underline{2[i,0,1,0,k-14-2i]}}\\
 &\overset{\left[(k-16)/2\right]}{\underset{i=0}{\oplus}}{\color{red}\overline{[i,0,2,0,k-16-2i]}} \\
 &\overset{\left[(k-13)/2\right]}{\underset{i=0}{\oplus}}{\color{red}\overline{[i,0,1,1,k-13-2i]}}
 \overset{\left[(k-17)/2\right]}{\underset{i=0}{\oplus}}{\color{blue}\underline{[i,0,1,1,k-17-2i]}} \\
 &\oplus {\color{red}\overline{[0,1,0,0,k-12]}} \overset{\left[(k-13)/2\right]}{\underset{i=1}{\oplus}}{\color{red}\overline{2[i,1,0,0,k-12-2i]}}
 {\underset{k-even}{\oplus}}{\color{red}\overline{[\frac{k-12}{2},1,0,0,0]}} \\
 &\overset{\left[(k-17)/2\right]}{\underset{i=0}{\oplus}}{\color{blue}\underline{2[i,1,0,0,k-16-2i]}}  {\underset{k-even}{\oplus}}{\color{blue}\underline{[\frac{k-16}{2},1,0,0,0]}} \\
 &\overset{\left[(k-16)/2\right]}{\underset{i=0}{\oplus}}{\color{red}\overline{[i,2,0,0,k-16-2i]}} \\
 &\overset{\left[(k-15)/2\right]}{\underset{i=0}{\oplus}}{\color{red}\overline{[i,1,0,1,k-15-2i]}} \overset{\left[(k-15)/2\right]}{\underset{i=0}{\oplus}}{\color{blue}\underline{[i,1,0,1,k-15-2i]}}\\
 &\overset{\left[(k-14)/2\right]}{\underset{i=0}{\oplus}}{\color{red}\overline{[i,1,1,0,k-14-2i]}} \overset{\left[(k-18)/2\right]}{\underset{i=0}{\oplus}}{\color{blue}\underline{[i,1,1,0,k-18-2i]}},  
\end{split}
\eeq
\beq
\begin{split}
\Sym^{k-14}(S)\otimes \wedge^7V &=
\overset{\left[(k-17)/2\right]}{\underset{i=0}{\oplus}}{\color{blue}\underline{[i,0,0,0,k-16-2i]}} \overset{\left[(k-13)/2\right]}{\underset{i=1}{\oplus}}{\color{red}\overline{[i,0,0,0,k-12-2i]}} \\
&\overset{\left[(k-16)/2\right]}{\underset{i=0}{\oplus}}{\color{red}\overline{2[i,0,0,1,k-15-2i]}} {\underset{k-odd}{\oplus}} {\color{red}\overline{[\frac{k-15}{2},0,0,1,0]}}\\
&\overset{\left[(k-15)/2\right]}{\underset{i=1}{\oplus}}{\color{blue}\underline{[i,0,0,1,k-15-2i]}} \\
&\overset{\left[(k-18)/2\right]}{\underset{i=0}{\oplus}}{\color{blue}\underline{2[i,0,1,0,k-18-2i]}} \oplus {\color{red}\overline{[0,0,1,0,k-14]}}  \\
&\overset{\left[(k-14)/2\right]}{\underset{i=1}{\oplus}}{\color{red}\overline{2[i,0,1,0,k-14-2i]}} \overset{\left[(k-17)/2\right]}{\underset{i=0}{\oplus}}{\color{red}\overline{[i,0,1,1,k-17-2i]}}\\
&\overset{\left[(k-16)/2\right]}{\underset{i=0}{\oplus}}{\color{blue}\underline{[i,1,0,0,k-16-2i]}} \overset{\left[(k-17)/2\right]}{\underset{i=0}{\oplus}}{\color{red}\overline{2[i,1,0,0,k-16-2i]}} \\
&{\underset{k-even}{\oplus}} {\color{red}\overline{[[\frac{k-16}{2}],1,0,0,0]}} \overset{\left[\frac{k-15}{2}\right]}{\underset{i=0}{\oplus}}{\color{red}\overline{[i,1,0,1,k-15-2i]}}\\
&\overset{\left[\frac{k-19}{2}\right]}{\underset{i=0}{\oplus}}{\color{blue}\underline{[i,1,0,1,k-19-2i]}} \overset{\left[\frac{k-18}{2}\right]}{\underset{i=0}{\oplus}}{\color{red}\overline{[i,1,1,0,k-18-2i]}},
\end{split}
\eeq
\beq
\begin{split}
\Sym^{k-16}(S)\otimes \wedge^8V &=
\overset{\left[(k-17)/2\right]}{\underset{i=0}{\oplus}}{\color{red}\overline{[i,0,0,0,k-16-2i]}} \overset{\left[(k-16)/2\right]}{\underset{i=1}{\oplus}}{\color{blue}\underline{[i,0,0,0,k-16-2i]}}\\
&\overset{\left[(k-15)/2\right]}{\underset{i=1}{\oplus}}{\color{red}\overline{[i,0,0,1,k-15-2i]}} \overset{\left[(k-19)/2\right]}{\underset{i=0}{\oplus}}{\color{blue}\underline{[i,0,0,1,k-19-2i]}} \\
&\overset{\left[(k-18)/2\right]}{\underset{i=0}{\oplus}}{\color{red}\overline{2[i,0,1,0,k-18-2i]}} \overset{\left[(k-20)/2\right]}{\underset{i=0}{\oplus}}{\color{blue}\underline{[i,1,0,0,k-20-2i]}} \\
&\overset{\left[(k-16)/2\right]}{\underset{i=0}{\oplus}}{\color{red}\overline{[i,1,0,0,k-16-2i]}} \overset{\left[(k-19)/2\right]}{\underset{i=0}{\oplus}}{\color{red}\overline{[i,1,0,1,k-19-2i]}},
\end{split}
\eeq
\beq
\begin{split}
\Sym^{k-18}(S)\otimes \wedge^{9}V&=
 \overset{\left[(k-16)/2\right]}{\underset{i=1}{\oplus}}{\color{red}\overline{[i,0,0,0,k-16-2i]}} \overset{\left[(k-20)/2\right]}{\underset{i=0}{\oplus}}{\color{blue}\underline{[i,0,0,0,k-20-2i]}} \\
 &\overset{\left[(k-19)/2\right]}{\underset{i=0}{\oplus}}{\color{red}\overline{[i,0,0,1,k-19-2i]}} \overset{\left[(k-20)/2\right]}{\underset{i=0}{\oplus}}{\color{red}\overline{[i,1,0,0,k-20-2i]}},
\end{split}
\eeq
\beq
\Sym^{k-20}(S)\otimes \wedge^{10}V=
\overset{\left[(k-20)/2\right]}{\underset{i=0}{\oplus}}{\color{red}\overline{[i,0,0,0,k-20-2i]}}
\eeq

The decompositions [Eqs.\ref{SymkS_10D}-\ref{SymkS5V_10D}] can be verified by dimension check. The dimensions of $\Sym^m\otimes \wedge^n V$ are given by the formula
\beqr
\dim(\Sym^m S\otimes\wedge^n V)&=& \dim(\Sym^m S)\dim(\wedge^n V)\nonumber\\
&=&\dbinom{s-1+m}{s-1}\dbinom{v}{n}=C^{s-1}_{s-1+m}C^{n}_{v} \label{dim SxV}
\eeqr
where $\dim(S)=s$, $\dim(V)=v$. The dimensions  of the RHS  can be obtained from Weyl dimension formula.  One can check that the RHS is a subrepresentation of the LHS, together with the dimension check this gives a rigorous proof of  [Eqs.\ref{SymkS_10D}-\ref{SymkS5V_10D}].

By the Schur's lemma an intertwiner between irreducible representations (a homomorphism of simple modules) is either zero or an isomorphism. This means that an intertwiner between non-equivalent
irreducible representations  always vanishes. This observation permits us to calculate the contribution of every irreducible representation to the cohomology separately.

Let us fix an irreducible representation $A$ and the number $k$. We will denote by $\nu _n$ (or by $\nu_n^k$ if it is necessary to show the dependence of $k$) the multiplicity of $A$ in $E^{kn}=\Sym^{k-2n}S\otimes \Lambda ^n V.$ The multiplicity of $A$ in the image of
$d:E^{kn}\to E^{k,{n-1}}$ will be denoted by $\kappa _n$, then the multiplicity of $A$ in the kernel of this map is equal to $\nu _n-\kappa _n$ and the multiplicity of $A$ in the cohomology $H^{kn}$
is equal to $h_n=\nu _n-\kappa _n-\kappa _{n+1}.$ It follows immediately that the multiplicity of $A$ in virtual representation $\sum _n (-1)^n H^{kn}$ (in the Euler characteristic) is equal to $\sum
(-1)^n\nu _n.$  It does not depend on $\kappa _n$, however, to calculate the cohomology completely we should know $\kappa _n$.  In many cases a heuristic calculation of cohomology can be based on a
principle that kernel should be as small as possible; in other words, the image should be as large as possible (this is an analog of the general rule of the physics of elementary particles: Everything
happens unless it is forbidden). In \cite{Bengt.Nilsson} this is called the principle of maximal propagation. {\footnote { Notice that the principle of maximal propagation should be applied to the composition of cohomology into irreducible representations of the full automorphism group.}} Let us illustrate this principle in the case when $k=9$ and $A=[0,1,0,0,1]$. In this case $\nu_4=1$, $\nu_3=3$, $\nu_2=1$. If we believe in the maximal propagation, then $\kappa_3=1,\kappa_4=1$, thus we have $\nu_3 - \kappa_3 - \kappa_4 =1$, and $[0,1,0,0,1] $ contributes only to  $H^{9,3}$.

In decompositions   [Eqs.\ref{SymkS_10D}-\ref{SymkS5V_10D}]  some terms are printed in red and overlined, some terms are printed in blue and underlined. We will prove that
 overlined red terms ${\color{red}\overline{[a,b,c,d,e]}}$  in $\Sym^mS\otimes \wedge^n V$  denoted later by  $B_n(k)$ where $k=m+2n$ are in the boundary (in  the image of the $n$-th differential $d_n$ in the cochain complex (Eq.~\ref{complex.10D}). The underlined blue terms ${\color{blue}\underline{[a,b,c,d,e]}}$  are mapped onto the boundary terms by the action of differential.  Both underlined and overlined terms do not contribute to cohomology.

These statements follow from  the maximal propagation principle, however in our situation we can give a rigorous proof of these statements by induction with respect to $k=m+2n$.

Let us assume that our statements are true for indices $<k$; in particular, $ B_n(k-1)$ consists of boundaries. We should prove that $B_n(k)$ also consists of boundaries. We will use the fact  that
 the differential $d$ commutes with multiplication by a polynomial depending on $t^{\alpha}$. To obtain the image of  $B_n(k-1)$ by multiplication by linear polynomial we should calculate $S\otimes B_n(k-1)$ and symmetrize with respect to the variables $t^{\alpha}$.

Generally, the tensor product of $S$ and a representation $[i,j,p,q,e]$ is given by the formula
\beq
\begin{split}
S\otimes [a,b,c,d,e]&=[a,b,c,d,e+1]+[a+1,b,c,d,e-1]\\
&+[a-1,b,c,d+1,e]+[a-1,b,c+1,d-1,e]\\
&+[a-1,b+1,c-1,d,e+1]+[a-1,b+1,c,d,e-1]\\
&+[a,b-1,c,d,e+1]+[a,b-1,c+1,d,e-1]\\
&+[a,b,c-1,d+1,e]+[a,b,c,d-1,e]\\
&+[a,b,c+1,d,e-1]+[a,b+1,c-1,d+1,e]\\
&+[a,b+1,c,d-1,e]+[a+1,b-1,c,d+1,e]\\
&+[a+1,b-1,c+1,d-1,e]+[a+1,b,c-1,d,e+1] \label{SxVector10D}
\end{split}
\eeq
To derive  (\ref {SxVector10D}) and (\ref {SxVector6D}) one can use the general result of \cite {BZ} giving an expression
of multiplicity $c_{\lambda, \nu} ^{\mu}$ of representation with highest weight $\mu$ in tensor product of representations with highest weights $\lambda$ and $\nu$ in terms of number of integral points in a polytope.  It follows from this general result that  for $\nu \gg 0$ the multiplicity $c_{\lambda, \nu }^{\mu}$ depends only of the difference $\mu -\nu$, therefore  checking (\ref {SxVector10D}) for finite number of cases we obtain a rigorous proof of it. We  used this idea with assistance of LiE code \cite {LiEcode}.
Using (\ref {SxVector10D}) and ( \ref {SymkS_10D}) we can describe the homomorphism
$S\otimes \Sym^{k-1}S\to \Sym ^kS$ and (multiplying by   $\Lambda ^nV$)  the homomorphism
$S\otimes E^{k-1,n} \to E^{kn}.$

It follows from this description that  all elements of $B_n(k)$ are boundaries if $B_n(k-1)$ consists of boundaries .
 Using this fact  one can derive the maximal propagation for $k$ from maximal propagation for $k-1$.

Let us consider as an example $A=[0,0,0,0,0]$, the scalar representation, for arbitrary $k$.  For all $k\neq 4,12$, we have $\nu_i=0$. For $k=4$, we have all $\nu_i$ vanish except $\nu_1=1$, hence all $\kappa_i$ vanish. The multiplicity of $[0,0,0,0,0]$ in $H^{4,1}$ is equal to $1$, and other cohomology $H^{4,i}$ do not contain scalar representation. For $k=12$, all $\nu_i$ vanish except $\nu_5=1$, hence $H^{12,5}$ contains $[0,0,0,0,0]$ with multiplicity $1$, and $H^{12,i}$ do not contain $[0,0,0,0,0]$ for $i\neq 5$. This agrees with Eq.~\ref{H4,1} and Eq.~\ref{Hk,5}, respectively.

\section {Calculations for D=6}

For $D=6$ spacetime, we have the cochain complex
\beq
\begin{split}
0 &\xleftarrow{d_0} \Sym^k(S) \xleftarrow{d_1} \Sym^{k-2}(S)\otimes V \xleftarrow{d_2} \Sym^{k-4}(S)\otimes \wedge^2 V \xleftarrow{d_3} \Sym^{k-6}(S)\otimes \wedge^3 V \nonumber \\
&\xleftarrow{d_4} \Sym^{k-8}(S)\otimes \wedge^4V \xleftarrow{d_5}\Sym^{k-10}(S)\otimes \wedge^5V \xleftarrow{d_6}\Sym^{k-12}(S)\otimes \wedge^6V \xleftarrow{d_7} 0 \label{complex.6D}
\end{split}
\eeq
where for $\Sym^m(S)\otimes \wedge^n(V)$, a grading degree defined by $k=m+2n$ is invariant upon homological differentials. All components of this complex can be regarded as representations of $\sso_6\times \sl_2$. We have
\beq
\begin{split}
&S=[0,0,1,1], \quad V=[1,0,0,0] \\
&\wedge^2V=[0,1,1,0], \quad \wedge^3V=[0,0,2,0]+[0,2,0,0],    \\
&\wedge^4V=\wedge^2V, \quad \wedge^5V=V, \quad \wedge^{6}V=[0,0,0,0],
\end{split}
\eeq

For $\Sym^m S\otimes \wedge^n V$, where $m\geq 1$,
\beq
\begin{split}
\Sym^k(S)&=
\overset{\lfloor\frac{k}{2}\rfloor}{\underset{i=1}{\oplus}}{\color{blue}\underline{[i,0,k-2i,k-2i]}}\oplus [0,0,k,k],  \label{SymkS_6D}
\end{split}
\eeq
\beq
\begin{split}
\Sym^{k-2}(S)\otimes V&=
\overset{\lfloor\frac{k}{2}\rfloor}{\underset{i=1}{\oplus}}{\color{red}\overline{[i,0,k-2i,k-2i]}}
\overset{\lfloor\frac{k-4}{2}\rfloor}{\underset{i=0}{\oplus}}{\color{blue}\underline{[i,0,k-4-2i,k-4-2i]}}\\
&\overset{\lfloor\frac{k-4}{2}\rfloor}{\underset{i=0}{\oplus}}{\color{blue}\underline{[i,1,k-3-2i,k-4-2i]}}
\overset{\lfloor\frac{k-3}{2}\rfloor}{\underset{i=1}{\oplus}}{\color{blue}\underline{[i,1,k-3-2i,k-2-2i]}}\\
&\oplus [0,1,k-3,k-2]
\end{split}
\eeq
\beq
\begin{split}
\Sym^{k-4}(S)\otimes \wedge^2V &=
\overset{\lfloor\frac{k-4}{2}\rfloor}{\underset{i=0}{\oplus}}{\color{red}\overline{[i,0,k-4-2i,k-4-2i]}}
\overset{\lfloor\frac{k-5}{2}\rfloor}{\underset{i=1}{\oplus}}{\color{blue}\underline{[i,0,k-4-2i,k-4-2i]}}\\
&\overset{\lfloor\frac{k-6}{2}\rfloor}{\underset{i=0}{\oplus}}{\color{blue}\underline{[i,0,k-4-2i,k-6-2i]}}\\
&\overset{\lfloor\frac{k-4}{2}\rfloor}{\underset{i=2}{\oplus}}{\color{blue}\underline{[i,0,k-4-2i,k-2-2i]}}\oplus[1,0,k-6,k-4]\\
&\overset{\lfloor\frac{k-8}{2}\rfloor}{\underset{i=0}{\oplus}}{\color{blue}\underline{[i,1,k-7-2i,k-8-2i]}}
\overset{\lfloor\frac{k-7}{2}\rfloor}{\underset{i=0}{\oplus}}{\color{blue}\underline{[i,1,k-7-2i,k-6-2i]}}\\
&\overset{\lfloor\frac{k-4}{2}\rfloor}{\underset{i=0}{\oplus}}{\color{red}\overline{[i,1,k-3-2i,k-4-2i]}}
\overset{\lfloor\frac{k-3}{2}\rfloor}{\underset{i=1}{\oplus}}{\color{red}\overline{[i,1,k-3-2i,k-2-2i]}}\\
&\overset{\lfloor\frac{k-6}{2}\rfloor}{\underset{i=0}{\oplus}}{\color{blue}\underline{[i,2,k-6-2i,k-6-2i]}}
\end{split}
\eeq
\beq
\begin{split}
\Sym^{k-6}(S)\otimes \wedge^3V &=
\overset{\lfloor\frac{k-5}{2}\rfloor}{\underset{i=1}{\oplus}}{\color{red}\overline{[i,0,k-4-2i,k-4-2i]}}
\overset{\lfloor\frac{k-6}{2}\rfloor}{\underset{i=0}{\oplus}}{\color{red}\overline{[i,0,k-4-2i,k-6-2i]}}\\
&\overset{\lfloor\frac{k-4}{2}\rfloor}{\underset{i=2}{\oplus}}{\color{red}\overline{[i,0,k-4-2i,k-2-2i]}}
\overset{\lfloor\frac{k-9}{2}\rfloor}{\underset{i=0}{\oplus}}{\color{blue}\underline{[i,0,k-8-2i,k-8-2i]}}\\
&\overset{\lfloor\frac{k-10}{2}\rfloor}{\underset{i=0}{\oplus}}{\color{blue}\underline{[i,0,k-8-2i,k-10-2i]}} \\
&\overset{\lfloor\frac{k-8}{2}\rfloor}{\underset{i=1}{\oplus}}{\color{blue}\underline{[i,0,k-8-2i,k-6-2i]}}\oplus [0,0,k-8,k-6]\\
&\overset{\lfloor\frac{k-8}{2}\rfloor}{\underset{i=0}{\oplus}}{\color{red}\overline{[i,1,k-7-2i,k-8-2i]}}
\overset{\lfloor\frac{k-8}{2}\rfloor}{\underset{i=0}{\oplus}}{\color{blue}\underline{[i,1,k-7-2i,k-8-2i]}}\\
&\overset{\lfloor\frac{k-7}{2}\rfloor}{\underset{i=0}{\oplus}}{\color{red}\overline{[i,1,k-7-2i,k-6-2i]}}
\overset{\lfloor\frac{k-7}{2}\rfloor}{\underset{i=1}{\oplus}}{\color{blue}\underline{[i,1,k-7-2i,k-6-2i]}}\\
&\overset{\lfloor\frac{k-10}{2}\rfloor}{\underset{i=0}{\oplus}}{\color{blue}\underline{[i,2,k-10-2i,k-10-2i]}}
\overset{\lfloor\frac{k-6}{2}\rfloor}{\underset{i=0}{\oplus}}{\color{red}\overline{[i,2,k-6-2i,k-6-2i]}} \label{SymkS3V_6D}
\end{split}
\eeq
\beq
\begin{split}
\Sym^{k-8}(S)\otimes \wedge^4V &=
\overset{\lfloor\frac{k-9}{2}\rfloor}{\underset{i=0}{\oplus}}{\color{red}\overline{[i,0,k-8-2i,k-8-2i]}}
\overset{\lfloor\frac{k-8}{2}\rfloor}{\underset{i=1}{\oplus}}{\color{blue}\underline{[i,0,k-8-2i,k-8-2i]}}\\
&\overset{\lfloor\frac{k-10}{2}\rfloor}{\underset{i=0}{\oplus}}{\color{red}\overline{[i,0,k-8-2i,k-10-2i]}}
\overset{\lfloor\frac{k-8}{2}\rfloor}{\underset{i=1}{\oplus}}{\color{red}\overline{[i,0,k-8-2i,k-6-2i]}}\\
&\overset{\lfloor\frac{k-12}{2}\rfloor}{\underset{i=0}{\oplus}}{\color{blue}\underline{[i,1,k-11-2i,k-12-2i]}}
\overset{\lfloor\frac{k-11}{2}\rfloor}{\underset{i=0}{\oplus}}{\color{blue}\underline{[i,1,k-11-2i,k-10-2i]}}\\
&\overset{\lfloor\frac{k-8}{2}\rfloor}{\underset{i=0}{\oplus}}{\color{red}\overline{[i,1,k-7-2i,k-8-2i]}}
\overset{\lfloor\frac{k-7}{2}\rfloor}{\underset{i=1}{\oplus}}{\color{red}\overline{[i,1,k-7-2i,k-6-2i]}}\\
&\overset{\lfloor\frac{k-10}{2}\rfloor}{\underset{i=0}{\oplus}}{\color{red}\overline{[i,2,k-10-2i,k-10-2i]}}
\end{split}
\eeq
\beq
\begin{split}
\Sym^{k-10}(S)\otimes \wedge^5 V&=
\overset{\lfloor\frac{k-8}{2}\rfloor}{\underset{i=1}{\oplus}}{\color{red}\overline{[i,0,k-8-2i,k-8-2i]}}
\overset{\lfloor\frac{k-12}{2}\rfloor}{\underset{i=0}{\oplus}}{\color{blue}\underline{[i,0,k-12-2i,k-12-2i]}}\\
&\overset{\lfloor\frac{k-12}{2}\rfloor}{\underset{i=0}{\oplus}}{\color{red}\overline{[i,1,k-11-2i,k-12-2i]}}
\overset{\lfloor\frac{k-11}{2}\rfloor}{\underset{i=0}{\oplus}}{\color{red}\overline{[i,1,k-11-2i,k-10-2i]}}\\
\end{split}
\eeq
\beq
\begin{split}
\Sym^{k-12}(S)\wedge^6 V&=
\overset{\lfloor\frac{k-12}{2}\rfloor}{\underset{i=0}{\oplus}}{\color{red}\overline{[i,0,k-12-2i,k-12-2i]}}
\end{split}
\eeq
As in the case $D=10$ the decompositions [Eqs.\ref{SymkS_6D}-\ref{SymkS3V_6D}] can be verified by dimension check.

Again we can prove that the terms printed in red and overlined (we denote them  by  $B_n(k)$ where $k=m+2n$)  are in the boundary and the terms  printed in blue and underlined  are mapped onto the boundary terms by the action of differential.  Both underlined and overlined terms do not contribute to cohomology.

One can derive these statements  from  the maximal propagation principle or give a rigorous proof by induction with respect to $k=m+2n$. To give the proof we use the formula for the tensor product of $S$ and a representation $[i,j,p,q]$ :
\beq
\begin{split}
S\otimes [i,j,p,q]&=[i,j,p+1,q+1]+[i+1,j,p-1,q-1]\\
&+[i,j-1,p,q-1]+[i,j-1,p,q+1]+[i,j,p+1,q-1]\\
&+[i+1,j,p-1,q+1]+[i-1,j+1,p,q-1]+[i-1,j+1,p,q+1] \label{SxVector6D}
\end{split}
\eeq
This formula allows us to compute the map $S\otimes E^{k-1,n}\to E^{k,n}$ transforming boundaries into boundaries.
One can prove using this map that  all elements of $B_n(k)$ are boundaries assuming that this is true for $B_n(k-1)$.

\section{Homology of super Poincare Lie algebra}

The super Poincare Lie algebra can be defined as super Lie algebra spanned by supersymmetry Lie algebra and Lie algebra of its group of automorphisms. {\footnote {Instead of Lie algebra of automorphisms one can take its subalgebra. For example, we can take as a subalgebra the orthogonal Lie algebra }}

  To calculate the homology and cohomology of super Poincare Lie algebra we will use the following statement proved by Hochschild and Serre \cite{HochSerre} .(It follows from
  Hochschild-Serre spectral sequence constructed in the same paper.)

Let $\cal P$ denote a Lie algebra represented as a vector space as a direct sum of two subspaces $\cal L$ and $\cal G$. We assume that $\cal G$ is an ideal in $\cal P$ and that $\cal L$ is semisimple.
It follows from the assumption that $\cal G$ is an ideal that $\cal L$ acts on $\cal G$ and therefore on cohomology of $\cal G$; the $\cal L$-invariant part of cohomology $H^{\bullet}(\cal G))$ will
be denoted by $H^{\bullet}(\cal G))^{\cal L}$. One can prove that
$$H^n({\cal P})=\sum_{p+q=n}H^p({\cal L})\otimes H^q(\cal G)^{\cal L}.$$

This statement remains correct if $\cal P$ is a super Lie algebra. We will apply it to the case when $\cal P$ is super Poincare Lie algebra, $\cal  G$ is the Lie algebra of supersymmetries and $L$ is the Lie algebra of automorphisms or its semisimple subalgebra.
(We are working with complex Lie algebras, but we can work with their real forms. The results do not change.)

Notice that it is easy to calculate the cohomology of semisimple Lie algebra $\cal L$; they are described by antisymmetric tensors on $\cal L$ that are invariant with respect to adjoint
representation. One can say also that they coincide with de Rham cohomology of corresponding compact Lie group. For ten-dimensional case $L=\sso_{10}$ and the compact Lie group is $\SO(10, {\bf R})$. Its
cohomology is a Grassmann algebra with generators of dimension 3,7,11,13 and 9. In general the cohomology of the group $\SO(2r, {\bf R})$ is a Grassmann algebra with generators $e_k$ having dimension
$4k-1$ for $k<r$ and the dimension $2r-1$ for $k=r$. The cohomology of Lie algebra $\sl_n$ coincide with the cohomology of compact Lie group $\SU(n)$; they form a Grassmann algebra with
generators of dimension $3, 5,...,2n-1.$

As we have seen only $\cal L$-invariant part of cohomology of Lie algebra of supersymmetries
contributes to the cohomology of super Poincare algebra. For $D=10$ this means that the only contribution comes from $(m,n)=(0,0), (m,n)=(2,1)$ and $(m,n)=(2,5)$, for $D=6$ the only contribution comes from  $(m,n)=(0,0)$ and $(m,n)=(2,1).$ (Here $m$ denotes the grading with respect to even ghosts $t^{\alpha}$ and $n$ the grading with respect to odd ghosts $c_m.$)

\end {document}